\documentclass[useAMS,usenatbib]{mn2e}
\usepackage{graphicx}                                            
\usepackage{times}
\usepackage{float}
\usepackage{rotating}
\usepackage{url}

\newcommand{\he}[1] {He\,{\sc #1}}

\newcommand{\ha}{H$\alpha$}

\def\kms{\mbox{${\rm km}\:{\rm s}^{-1}\:$}}

\def\lesssim{\mathrel{\hbox{\rlap{\hbox{\lower4pt\hbox{$\sim$}}}\hbox{$<$}}}}
\let\la=\lesssim
\def\gtrsim{\mathrel{\hbox{\rlap{\hbox{\lower4pt\hbox{$\sim$}}}\hbox{$>$}}}}
\let\ga=\gtrsim

\def\lx{$L_\mathrm{X}$}

\def\ergs{erg s$^{-1}$}

\def\sloanu{$u^\prime$}
\def\sloang{$g^\prime$}
\def\sloanr{$r^\prime$}
\def\sloani{$i^\prime$}
\def\sloanz{$z^\prime$}
\title[The optical counterpart of Swift J1745-26 ]{The optical counterpart of the bright X-ray transient Swift J1745-26}
\author[Mu\~noz-Darias et al.]{T.~Mu\~noz-Darias$^{1}$, A. de Ugarte Postigo$^{2,3}$, D. M. Russell$^{4,5}$,  S.~Guziy$^{6,2}$, J. Gorosabel$^{2,7,8}$,
\newauthor
J. Casares$^{4,5}$, M.~ Armas Padilla$^{9}$, P.~A.~Charles$^{1,10}$, R.~P.~Fender$^{1}$, T.~M.~Belloni$^{11}$, 
F.~Lewis$^{12}$, \newauthor
S.~Motta$^{13}$, A.~Castro-Tirado$^{2}$, C.~G.~Mundell$^{14}$, R.~S\'anchez-Ram\'irez$^{2}$, C.~C.~Th\"one$^{2}$\\
$^{1}$School of Physics and Astronomy University of Southampton, Southampton, Hampshire, SO17 1BJ, United Kingdom\\
$^{2}$ Instituto de Astrof\'{\i}sica de Andaluc\'{\i}a (IAA-CSIC), Glorieta
de la Astronom\'{\i}a s/n, 18008, Granada, Spain. \\
$^{3}$Dark Cosmology Centre, Niels Bohr Institute, Juliane Maries Vej 30, Copenhage \O, 2100, Denmark\\
$^{4}$Instituto de Astrof\'isica de Canarias (IAC), V\'ia L\'actea s/n, La Laguna, E-38205, S/C de Tenerife, Spain\\
$^{5}$Departamento de Astrof\'isica, Universidad de La Laguna, La Laguna, E-38205, S/C de Tenerife, Spain\\
$^{6}$ Nikolaev National University, Nikolska 24, Nikolaev, 54030, Ukraine\\
$^{7}$ Unidad Asociada Grupo Ciencia Planetarias UPV/EHU-IAA/CSIC, Departamento de F\'{\i}sica Aplicada I, E.T.S. Ingenier\'{\i}a,\\ Universidad del Pa\'{\i}s Vasco UPV/EHU, Alameda de Urquijo s/n, E-48013,  Bilbao, Spain.\\
$^{8}$ Ikerbasque, Basque Foundation for Science, Alameda de Urquijo 36-5, E-48008 Bilbao, Spain.\\
$^{9}$ Astronomical Institute 'Anton Pannekoek', University of Amsterdam, Science Park 904, 1098 XH, Amsterdam, the Netherlands\\
$^{10}$Department of Astronomy, University of Cape Town, Private Bag X3, Rondebosch 7701, Republic of South Africa\\
$^{11}$ INAF-Osservatorio Astronomico di Brera, Via E. Bianchi 46, I-23807 Merate (LC), Italy\\
$^{12}$Faulkes Telescope Project, University of Glamorgan, Pontypridd, CF37 1DL, UK\\
$^{13}$ European Space Astronomy Centre (ESAC)/ESA, PO Box 78, E-28691 Villanueva de la Ca\~nada, Madrid, Spain\\
$^{14}$ Astrophysics Research Institute, Liverpool John Moores University, Twelve Quays House, Egerton Wharf, Birkenhead, CH41 1LD, UK
}

\begin{document}
\maketitle

\begin{abstract}
We present a 30-day monitoring campaign of the optical counterpart of the bright X-ray transient Swift J1745-26, starting only 19 minutes after the discovery of the source.  We observe the system peaking at \sloani$ \sim 17.6$ on day 6 (MJD 56192) to then decay at a rate of $\sim 0.04$ mag day$^{-1}$. We show that the optical peak occurs at least 3 days later than the hard X-ray (15-50 keV) flux peak. Our measurements result in an outburst amplitude greater than 4.3 magnitudes, which favours an orbital period $\lesssim$ 21 h and a companion star with a spectral type later than $\sim$ A0. Spectroscopic observations taken with the GTC-10.4 m telescope reveal a broad (\textit{FWHM} $\sim 1100$ \kms), double-peaked \ha~ emission line from which we constrain the radial velocity semi-amplitude of the donor to be $K_2 > 250$ \kms. The breadth of the line and the observed optical and X-ray fluxes suggest that Swift J1745-26 is a new black hole candidate located closer than $\sim 7$ kpc.       

\end{abstract}
\section{Introduction}
Low mass X-ray binaries (LMXBs) are interacting binaries harbouring a neutron star (NS) or a black hole (BH) accreting from a companion star typically lighter than the Sun. Accretion takes place via an accretion disc, where gravitational energy is efficiently converted into radiation \citep{Shakura1973}. LMXBs are multiwavelength sources, emitting from high-energies to radio through different thermal and non-thermal processes (e.g. \citealt{Remillard2006b}, \citealt{Fender2006}). If the mass transfer rate is high enough, these systems are always bright, with X-ray luminosities in the range \lx$ \sim 10^{36-39}$ \ergs. They are so-called persistent sources, whereas LMXBs with lower mass transfer rates tend to be found as X-ray binary transients (XRTs). These objects spend most part of their lives in a dim, quiescent state, displaying luminosities as low as \lx$ \sim 10^{31}$ \ergs. However, with recurrence times of a few months to decades, they undergo periods of activity, becoming as bright as persistent systems. It is during these  outbursts, typically lasting a few weeks to months, when they are discovered by X-ray telescopes and subsequently studied using multiwavelength facilities.\\
Active X-ray binaries display a well known phenomenology at high energies (e.g. \citealt{vanderklis2006}; \citealt{Belloni2011}). They also show very distinctive optical features, such as broad double-peaked emission lines testifying to the presence of an accretion disc (e.g. \citealt{Charles2006}). Galactic BHs are mostly found as XRTs, whereas the vast majority of the persistent population harbour NSs. This is not well understood yet but could be related to the dependence on the compact object mass of the critical mass transfer rate for a system to be persistent (e.g. \citealt{King1996}).\\
Swift J1745-26 (Swift J174510.8-262411; hereby J1745)  was discovered by the Burst Alert Telescope (BAT; \citealt{Barthelmy2005}) on board the \textit{Swift X-ray observatory} on September 16, 2012 (Cummings et al. 2012a, 2012b). Subsequent observations performed by \textit{Swift} and \textit{INTEGRAL} at X-ray wavelengths showed a significant flux increase during the following days together with X-ray properties resembling those typically seen in BH transients (e.g. \citealt{Sbarufatti2012}, \citealt{Tomsick2012}, \citealt{Belloni2012}).  Multiwavelength observations supported the X-ray binary nature of the system, although a definitive evidence for a BH accretor has not been reported yet (see \citealt{Rau2012}, \citealt{deUgartePostigo2012}, \citealt{Russell2012a} for optical/infrared; see also \citealt{Miller-Jones2012}, \citealt{Corbel2012} for radio).\par 
Here, we present the first study of the optical counterpart of J1745. It includes spectroscopic data and multi-band photometric follow-up from a few minutes after its discovery until the source was no longer visible due to Sun constraints. We compare the outburst evolution in the optical, with that at high-energies (\textit{Swift}/BAT 15-50 keV ) and report the first constraints on some of the orbital parameters of the system.

\section{Observations and Results}
\label{sec:obs}

\begin{table}
\caption{Observing log.}             
\label{table:1}      
         
\begin{tabular}{ c c c c }     
\hline\hline       
MJD & Telescope & Band(s) & T$_{exp}$ (s) \\ 
\hline                    
 56186.406 & FTS   & \textit{V}, \textit{R}, \sloani & 200, 100, 100 \\
 56187.85 & GTC  & \sloani & 60 ($\times$ 10)\\
 56187.87 & GTC  & Spec. R1000R & 900 ($\times$ 2) \\
 56187.89 & GTC  & Spec. R1000B & 900 ($\times$ 2)\\
 
 56188.88 & IAC80 & I & 300 ($\times$ 8) \\
 56191.418 & FTS   & \textit{V}, \textit{R}, \sloani & 200, 100, 100 \\
 56191.817 & IAC80 & I & 30 ($\times$ 11) \\
 56195.422 & FTS   & \textit{V}, \sloani, \textit{R} & 200, 100, 100 \\
 56195.829 & LT & \sloani & 100 \\
 56198.825 & LT & \sloani & 100 \\
 56199.823 & LT & \sloani & 100 \\
 56201.827 & LT & \sloani & 100 \\
 56204.46 & FTS    & \textit{R}, \sloani & 100 ($\times$ 2), 100\\
 56204.877 & CAHA-2.2m & \sloanu, \sloang, \sloanr, \sloanz & 4$\times$(100, 100, 100, 100)\\
 56206.819 & LT & \sloani & 100 \\
 56207.819 & LT & \sloani & 100 \\
 56208.818 & LT & \sloani & 100 \\
 56209.817 & LT & \sloani & 100 \\
 56210.817 & LT & \sloani & 100 \\
 56211.816 & LT & \sloani & 100 \\
 56214.811 & LT & \sloani & 100 \\
 56215.812 & LT & \sloani & 100 \\
\hline                  
\end{tabular}
\end{table}

\begin{figure}
\centering
\includegraphics[height=6 cm,keepaspectratio]{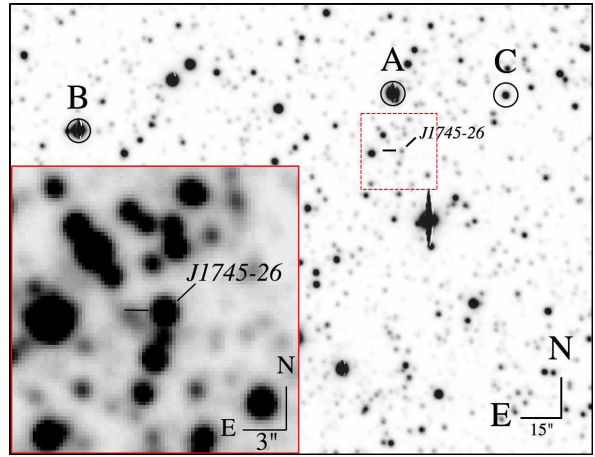}
\caption{Finding chart of Swift J1745-26 in Sloan \sloani~band taken with the GTC-10.4 m telescope. The refined object coordinates (J2000) are 17:45:10.85, -26:24:12.6, for which we estimate an absolute error of 0.3$^{~\prime\prime}$. The three comparison stars used to carry out the flux calibration are labelled as A (\sloani=13.4), B (\sloani=13.2) and C (\sloani=16.2).}
\label{fc}
\end{figure}

Our optical follow-up of J1745  began a mere 19 minutes after the \textit{Swift}/BAT alert and continued for $\sim$ 30 days. Five different facilities were utilized: (i) the 2 m Faulkes Telescope South (FTS; located at Siding Spring, Australia), (ii) OSIRIS attached to the \textit{Gran Telescopio de Canarias} (GTC) 10.4 m telescope in La Palma (Spain), (iii) the 2.0 m Liverpool Telescope (LT) also in La Palma, (iv) the IAC80 82 cm telescope in Tenerife (Spain) and (v) the 2.2 m telescope at Calar Alto observatory (CAHA-2.2m) in Almeria (Spain). An observing log is presented in Table \ref{table:1}. 
Observations were taken mostly using the Sloan \sloani filter, although \textit{V} and \textit{R} Johnson, I Bessel and Sloan \sloanu, \sloang, \sloanr, \sloanz~bands were also used for some epochs. Bias and flat-field corrections were performed using \textsc{iraf} routines and the flux of the optical counterpart was extracted and calibrated using three photometric comparison stars (Fig.  \ref{fc}). The error in our absolute calibration is $\sim$ 0.1 mags. However, the relative errors between data points are much smaller, allowing us a detailed study of the outburst evolution. The \sloani band light-curve is presented in Fig. \ref{lc}, for which, we also used I band data converted to \sloani using the transformations of \cite{Jordi2006}. In total, our photometry covers the outburst evolution of J1745 with 19 visits.\par
The source is observed to brighten during the first $\sim$ 6 days after the X-ray discovery at an average rate of $\sim 0.1$ mag day$^{-1}$, peaking at \sloani$\sim 17.6$ between days 5--10. Since day 10, a decay down to $\sim 18.1$ at a rate of $\sim 0.04$ mag day$^{-1}$ is observed. We note that given the quiescence level (\sloanr $> 23.1$) reported by \cite{Hynes2012}, we missed the major part of the outburst rise.  
Observations with filters other than \sloani were also taken on a few occasions (Tab. \ref{table:1}). Our best \textit{V} band determined magnitude is $20.0 \pm 0.2$ on MJD 56191, the other two epochs having larger errors and being consistent with the earlier one. On MJD 56204 using the CAHA-2.2m we measure \sloang=$21.29\pm0.12$, \sloanr=$19.14\pm0.04$ and \sloanz=$16.82\pm0.04$, whereas the source was not detected in \sloanu ($>$21). The same day we obtain \sloani=$17.96\pm0.03$ and \textit{R}=$18.45\pm0.04$ using the FTS. We observe the \textit{R}-\sloani colour to be $0.86\pm0.09$,  $0.92\pm0.08$ and $0.97\pm0.07$ on days MJD 56186, 56191 and 56195, respectively. Then, after the peak flux (and transition towards softer states; \citealt{Belloni2012}), it becomes bluer, being $0.80\pm0.05$ on MJD 56204. 

\subsection{Spectroscopy}
\begin{figure}
\centering
\includegraphics[width= 9 cm,height=6.5cm]{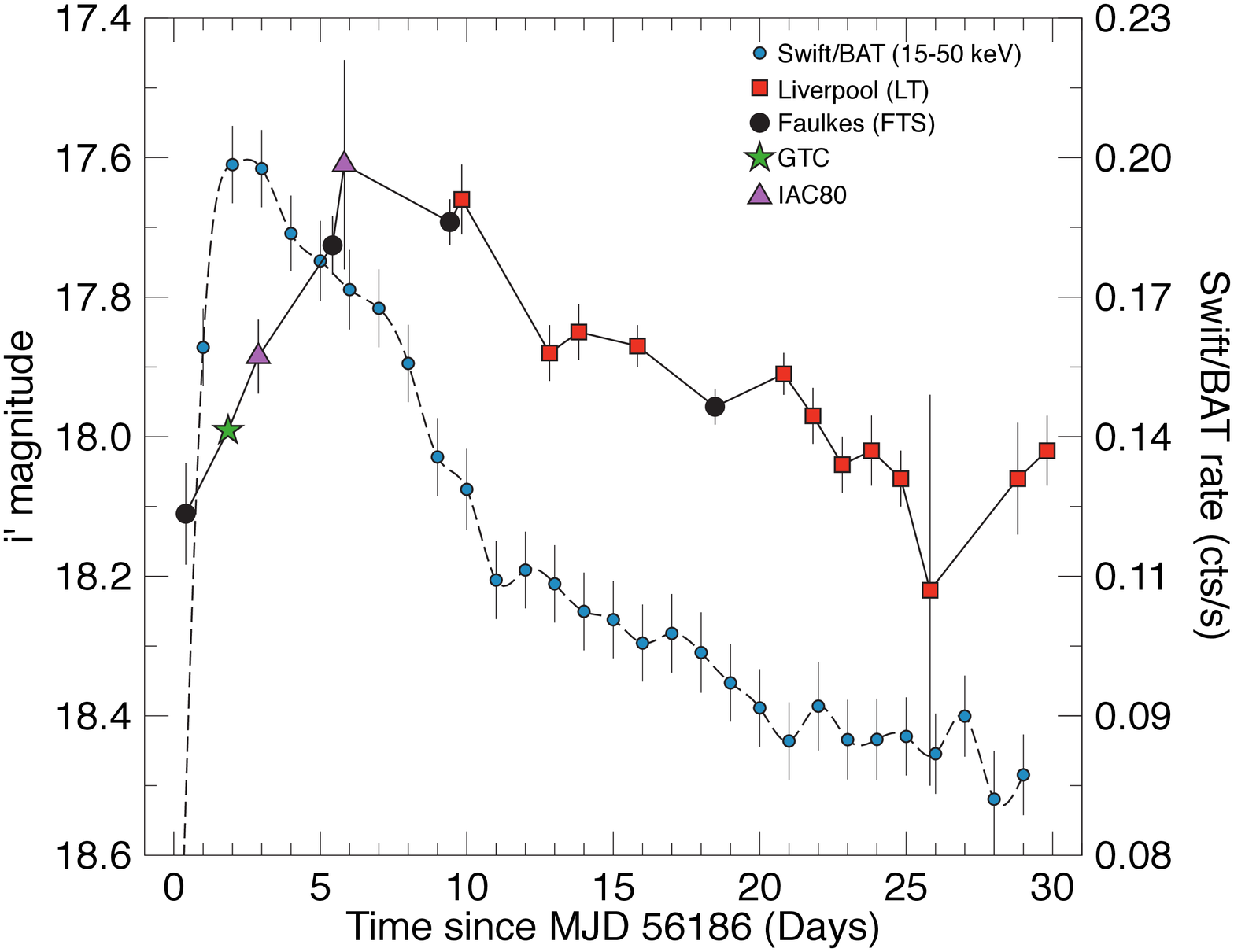}
\caption{Optical \sloani~light-curve of J1745 obtained by combining all the available photometry. The one-day average \textit{Swift}/BAT (15-50 keV) lightcurve (rescaled and offset) is shown as a dashed line.}
\label{lc}
\end{figure}

Four optical spectra were obtained on Sep 17, 2012 between 20:44:47 and 21:48:06 UT (i.e. only $\sim$ 35 hours after the discovery of the source) using the spectroscopic mode of OSIRIS/GTC. The system was at \sloani$ \sim 18$ magnitude at that time (see Fig. \ref{lc}). Observations consisted of $2 \times 900$s exposures using the R1000R grating, which covers the range from 5000 to 10000 {\AA} at a resolution power of $\lambda/\Delta\lambda \sim 1100$, followed by $2 \times 900$s using the R1000B grating, with a wavelength coverage from 3600 to 7500 {\AA} and a similar resolution. The slit was placed at the parallactic angle with a width of 1". Unfortunately, observations were performed at a necessarily high airmass ($>2.0$) and hence high extinction. This considerably limited our analysis, which was effectively restricted to the 5000--7500 \AA~ range covered in all the spectra. Data reduction was performed using standard IRAF procedures. No reliable flux calibration was possible due to the high airmass and variable conditions.\par
\begin{figure}
\centering
\includegraphics[width= 7.cm,height=6.5cm]{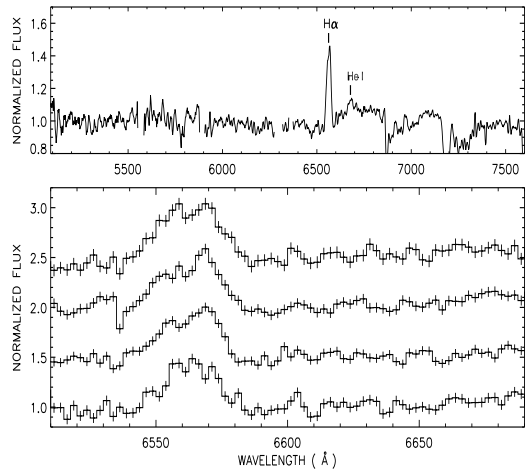}
\caption{Top panel: normalized, average spectrum of J1745. Gaps are due to significant residuals after the sky subtraction.  Bottom panel: zoom in of the  \ha~region. An offset of 0.5, 1. and 1.5 has been applied to spectra 2, 3 and 4, respectively.}
\label{halfa}
\end{figure}

The  spectrum is almost featureless in the 5000-7500 \AA~ range (top panel in Fig. \ref{halfa}), except for the presence of a broad \ha~emission line, which is clearly detected in the four individual spectra (see bottom panel in Fig. \ref{halfa}). This feature, typically observed in compact binaries, has a double-peaked, asymmetric profile, as expected from an accretion disc origin \citep{Smak1969}. The shape of the line, in particular the blue peak, is seen to vary from spectrum to spectrum along the $\sim 1$ hour of our continuous monitoring (i.e. on time  scales of $\sim 15$ min; ). A Gaussian fit to the average \ha~profile gives an equivalent width of $EW=12.6 \pm 0.5$~\AA~and \textit{FWHM}$=1115 \pm 38$~\kms, whereas a peak-to-peak separation of $634 \pm 18$ \kms is obtained by fitting the line with two Gaussians. These measurements are consistent with those found in other X-ray binaries and are discussed in section \ref{sec:discussion}. 
Finally, we note that the \he{i} $\lambda6678$ emission line, typically seeing in X-ray binaries, is also detected in the average spectrum of J1745 (top panel in Fig. \ref{halfa}).

\section{Discussion}
\label{sec:discussion}

We have studied the evolution of the optical counterpart of the X-ray transient Swift J1745-26 using photometry and spectroscopy. The spectrum is dominated by a broad (\textit{FWHM} $\sim 1100$ \kms), double-peaked \ha~emission. Since these lines are known to be naturally formed in geometrically thin accretion discs (\citealt{Smak1969}; \citealt{Horne1986}), the detection confirms both the association of the proposed optical counterpart with the X-ray source and its X-ray binary nature. Indeed, \ha~ is typically the most prominent optical emission line in X-ray binaries. Its \textit{EW} can be as large as $\sim 100$ \AA~ during the quiescence phase, becoming smaller ($\lesssim 20$ \AA) during outburst. Therefore, our measurement ($EW\sim13$~\AA) is consistent with typical values observed during the earliest phases of the outburst (\citealt{Fender2009}). 

Optical emission from active X-ray binaries arises in regions typically a few light seconds from the central object (see e.g. \citealt{Hynes2006}, \citealt{Munoz-Darias2007}). This is mainly a result of X-ray reprocessing in the outer accretion disc (\citealt{vanParadijs1994}) but with some possible synchrotron jet contribution during hard X-ray states \citep{Russell2006}. Assuming that the \ha~ emission originates in a Keplerian, outer disc rim, the broadness of its profile tells us about the projected velocity of the regions closer to its inner radius (see \citealt{Horne1986}). Therefore, it is expected to, at least, depend on (i) the mass of the compact object, (ii) the orbital inclination and (iii) the size of the accretion disc. Not surprisingly, the systems with broadest \ha~ emission lines are found to be black holes with relatively high inclination e.g. XTE J1118+480 (\textit{FWHM} $\sim 2500$ \kms; \citealt{Torres2004}). However, these measurements are taken in quiescence, where \textit{FWHM} tends to be larger. For instance, in the BH binary GRS~1009-45 (Nova Vela 1993) \textit{FWHM} $\sim 2000$ \kms~ is measured during quiescence (\citealt{Shahbaz1996}; \citealt{Filippenko1999}) and \textit{FWHM} $\sim 1370$ \kms~ in outburst (\citealt{DellaValle1993}). NS systems seem to follow the same trend, but displaying lower velocities. In quiescence, Cen X-4 (low inclination; \citealt{Shahbaz1993}) shows \textit{FWHM} $\sim 640$ \kms~  (\citealt{Torres2002}) and XTE~J2123-058 (grazing eclipses; \citealt{Zurita2000}) displays \textit{FWHM} $\sim 1300$ \kms~\citep{Casares2002}. For the latter, we estimate \textit{FWHM} $\sim$ 500 -- 600 \kms~ from the outburst spectrum shown in \cite{Hynes2001} and similar values have been observed in the eclipsing NS systems X1822-371 \citep{Harlaftis1997} and EXO 0748-676 \citep{Pearson2006} also during active phases. In light of the above, the \textit{FWHM} measured for J1745 fits better with a BH scenario. However, we note that the amount of measurements available is relatively low and that other orbital parameters like the orbital period (i.e. size of the disc) should play a role in this discussion. \par

\subsection{Outburst evolution}
The photometric follow-up presented here started very promptly after the \textit{Swift} alert. However, in our first measurement the system had already brightened considerably from its quiescent level \citep{Hynes2012}. During the first $\sim$ 6 days, it kept rising at a rate of $0.1$ mag day$^{-1}$ before peaking at \sloani=17.6 . This rate seems comparable with those seen in other XRTs. For instance, $0.36$ mag day$^{-1}$ was observed in XTE J1118+480 \citep{Zurita2006}, $0.14$ mag day$^{-1}$ in Aql X-1 \citep{Shahbaz1998a} and $\sim 0.25$ mag day$^{-1}$ in GRO J0422+32 \citep{Castro-Tirado1997}. After the peak, we see a decay at a rate of $\sim 0.04$ mag day$^{-1}$ also comparable to the 0.05 -- 0.07 mag day$^{-1}$ rate reported in the aforementioned works. At the end of our 1-month monitoring J1745 was still bright and far from quiescence.\par 

In Fig. \ref{lc} we compare the optical and (15--50 keV) X-ray light-curves directly. This clearly shows the X-ray peak occurring $\sim$ 3d before the optical. X-rays peaking before the optical emission is at odds with some observations of other XRTs, where the flux is seen to peak (and/or suggested to rise) earlier in the optical (e.g. \citealt{Orosz1997}, \citealt{Shahbaz1998a}, \citealt{Zurita2006}). This is traditionally interpreted as a proof of an \textit{outside-in} outburst propagation. However, we note that those works typically use softer bands (e.g. 2--10 keV) than that here (15--50 keV). If we consider the canonical outburst evolution (e.g. \citealt{Belloni2011}), soft X-rays will peak several days later than the BAT data. Indeed, preliminary work on this source by part of our team shows that the maximum of the soft X-ray emission occurs around the same date or even later than the optical peak we report in this work \footnote{See http://www.rssd.esa.int/SD/INTEGRAL/images/POM2/2013-01.jpg}.\par
\begin{figure}
\centering
\includegraphics[width= 6.5cm,height=8.cm,angle=-90]{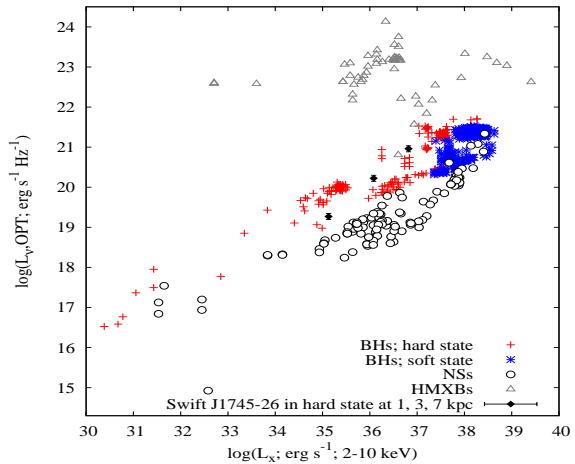}
\caption{Optical/X-ray luminosity diagram from Russell et al. (2006; 2007). Black holes (soft and hard states), neutron stars LMXBs and high mass x-ray binaries (HMXBs) are included in the plot. Swift J1745-26 is consistent with being a BH in the hard state for the 3 distances considered (from left to right 1, 3 and 7 kpc; see text).}
\label{diagram}
\end{figure}

 
Adopting the spectral parameters reported by \cite{Tomsick2012}, the peak X-ray flux corresponds to 4.5$\times 10^{-8}$ ergs cm$^{-2}$ s$^{-1}$ in the 15--50 keV band (2.9$\times 10^{-8}$ ergs cm$^{-2}$ s$^{-1}$ within 2-10 keV). This corresponds to $\sim 1.2$ Crab (2-10 keV), making J1745 one of the brightest XRTs in recent times. Assuming the accretion rate does not exceed the Eddington limit, the maximum distance to the source is $d \sim 7$ kpc for a BH and $\sim 3$ kpc for a NS\footnote{Here we extrapolate the observed flux to the 0.1--100 keV band ($\sim 1.4\times 10^{-7}$ ergs cm$^{-2}$ s$^{-1}$). We assume $1.4 M_{\odot}$ for a neutron star and the $8 M_{\odot}$ average mass for stellar mass black holes reported by \cite{Ozel2010}.}. After the BAT peak, J1745 is observed to move towards softer X-ray states (\citealt{Belloni2012}). Since BHs undergo state transition at luminosities higher than 2\% of the Eddington luminosity \citep{Maccarone2003}, we estimate $1\lesssim \frac{d}{1~\mathrm{kpc}}\lesssim 7$ for a BH accretor. Using these constraints on the distance, we have over plotted the X-ray flux reported by \cite{Sbarufatti2012} together with our optical measurement (MJD 56186; i.e hard X-ray state) in the optical--X-ray luminosity diagram presented by Russell et al. (2006; 2007) for $d = 1, 3$ and 7 kpc (Fig. \ref{diagram}). Here, we have used the the column density ($N_H=1.70 \pm 0.04 \times 10^{22}$ cm$^{-2}$) from \cite{Tomsick2012}. We find that regardless of the distance assumed, J1745 is consistent with being a BH in the hard state. However, we note that smaller $N_H$ values will result in the system being consistent with a nearby ($\sim 1$ kpc) NS LMXB. 

\subsection{Orbital parameters}
The results presented here can be used to constrain the orbital parameters of J1745. The \ha~peak-to-peak separation of $634 \pm 18$ \kms encodes information regarding the outer accretion disc velocity and was empirically related to the companion star projected velocity ($K_2$) by \cite{Orosz1994}. Defining the outer disc velocity ($v_d$) as half of the peak-to-peak separation (i.e. $v_d=317\pm9$~\kms for J1745), they find that $v_d/K_2\sim$1.1--1.25, which yields $K_2 > 250$ \kms for J1745. We note that this value is a lower limit, since the Orosz et al. relation is obtained from quiescent accretion discs, which have larger $v_d$ as result of a smaller outer disc radius than in outburst (see \citealt{Corral-Santana2013} for a discussion). \par 
Our photometric data show a peak \sloani~band magnitude of 17.6 corresponding to $\sim 0.3$ mJy. Assuming that the \sloanr-\sloani$\cong 1.2$ colour (MJD 56204) is the same on MJD 56195 we obtain an outburst amplitude in the SDSS-r band  $\Delta$ \sloanr$>4.3$ from the quiescent level reported by \cite{Hynes2012}. \cite{Shahbaz1998} found a relation between the outburst amplitude in V band ($\Delta V$) and the orbital period ($P_{orb}$). Using $\Delta V=\Delta$\sloanr we obtain $P_{orb}\lesssim 21$ hours. The last assumption seems reasonable since during the outburst the optical spectrum (once corrected from extinction) is expected to be disc dominated and relatively flat (\citealt{vanParadijs1994}) whereas a strong contribution from a redder companion star is expected in quiescence.
We note that in the case of $\Delta V > \Delta$\sloanr~the orbital period would be smaller. The Shahbaz \& Kuulkers relation is valid for orbital periods $P_{orb}\lesssim 1$~day, which seems consistent with the dim, quiescent optical counterpart. Periods longer than $\sim$ 1d would imply evolved companion stars with likely brighter quiescence levels (see e.g. \citealt{King1993}, \citealt{Munoz-Darias2008}). Using the relation reported in \cite{Faulkner1972}, our constraint on the orbital period results in a mean density for the donor star  $> 0.2$ g cm$^{-3}$. For a main sequence companion, this limit yields a spectral type later than $\sim$ A0 (\citealt{Cox2000}).
\section{Conclusions}        
We have undertaken a detailed follow-up of the optical counterpart of the bright X-ray transient Swift J1745-26. All the observables suggest that Swift J1745-26 is a new black hole candidate, as proposed by preliminary analysis of X-ray observations. We provide the first constraint on some of the fundamental parameters of this X-ray binary.
\begin{itemize}
\item The optical spectrum of J1745 shows a strong, double-peaked~\ha~emission line from which we infer a donor radial velocity semi-amplitude of $K_2 > 250$ \kms. We also show that the breadth of this line (\textit{FWHM} $\sim 1100$ \kms) suggests a black hole accretor.   
\item Our photometric campaign revealed an outburst amplitude $>$ 4.3 magnitudes, favouring an orbital period lower than $\sim$ 21 h and a companion star with a spectral type later than $\sim$ A0.
\item The optical emission peaks at least 3 days after the hard X-rays. The observed optical and X-ray fluxes are consistent with J1745 being a black hole in the hard state lying at a distance of $1\lesssim \frac{d}{1~\mathrm{kpc}}\lesssim 7$.      
\end{itemize}

\vspace{1cm}

\noindent We thank Antonio Cabrera-Lavers and Alison Tripp for contributing to the GTC and Faulkes observing programs. TMD and DMR acknowledge funding via an EU Marie Curie Intra-European Fellowships under contract numbers 2011-301355 and 2010-274805. AdUP acknowledges support by the EU Marie Curie Career Integration Grant FP7-PEOPLE-2012-CIG 322307. AdUP, CT and JG acknowledge support by Spanish research project AYA2012-39362-C02-02. JC acknowledges support by the Spanish research project AYA2010-18080. Partially funded by the Spanish Consolider-Ingenio 2010 Programme grant CSD2006-00070: First Science with the GTC and the EU Programme FP7/2007-2013 under grant agreement number ITN 215212 \textquotedblleft Black Hole Universe\textquotedblright. The Dark Cosmology Centre is funded by the DNRF. The Faulkes Telescope South is maintained and operated by Las Cumbres Observatory Global Telescope Network. This study was carried out in the framework of the Unidad Asociada IAA-CSIC at the group of planetary science of ETSI-UPV/EHU, and supported by the Ikerbasque Foundation for Science.

\bibliographystyle{mn2e}
\bibliography{/Users/tmd/Dropbox/Libreria.bib}  

\begin{thebibliography}{}

\bibitem[\protect\citeauthoryear{{Barthelmy}, {Barbier}, {Cummings},
  {Fenimore}, {Gehrels}, {Hullinger}, {Krimm}, {Markwardt}, {Palmer},
  {Parsons}, {Sato}, {Suzuki}, {Takahashi}, {Tashiro} \& {Tueller}}{{Barthelmy}
  et~al.}{2005}]{Barthelmy2005}
{Barthelmy} S.~D.,  {Barbier} L.~M.,  {Cummings} J.~R.,  {Fenimore} E.~E.,
  {Gehrels} N.,  {Hullinger} D.,  {Krimm} H.~A.,  {Markwardt} C.~B.,  {Palmer}
  D.~M.,  {Parsons} A.,  {Sato} G.,  {Suzuki} M.,  {Takahashi} T.,  {Tashiro}
  M.,    {Tueller} J.,  2005, \ssr, 120, 143

\bibitem[\protect\citeauthoryear{{Belloni}, {Cadolle Bel}, {Casella} \& {et
  al.}}{{Belloni} et~al.}{2012}]{Belloni2012}
{Belloni} T.,  {Cadolle Bel} M.,  {Casella} P.,    {et al.} 2012, The
  Astronomer's Telegram, 4450, 1

\bibitem[\protect\citeauthoryear{{Belloni}, {Motta} \&
  {Mu{\~n}oz-Darias}}{{Belloni} et~al.}{2011}]{Belloni2011}
{Belloni} T.~M.,  {Motta} S.~E.,    {Mu{\~n}oz-Darias} T.,  2011, Bulletin of
  the Astronomical Society of India, 39, 409

\bibitem[\protect\citeauthoryear{{Casares}, {Dubus}, {Shahbaz}, {Zurita} \&
  {Charles}}{{Casares} et~al.}{2002}]{Casares2002}
{Casares} J.,  {Dubus} G.,  {Shahbaz} T.,  {Zurita} C.,    {Charles} P.~A.,
  2002, \mnras, 329, 29

\bibitem[\protect\citeauthoryear{{Castro-Tirado}, {Ortiz} \&
  {Gallego}}{{Castro-Tirado} et~al.}{1997}]{Castro-Tirado1997}
{Castro-Tirado} A.~J.,  {Ortiz} J.~L.,    {Gallego} J.,  1997, \aap, 322, 507

\bibitem[\protect\citeauthoryear{{Charles} \& {Coe}}{{Charles} \&
  {Coe}}{2006}]{Charles2006}
{Charles} P.~A.,  {Coe} M.~J.,  2006, in Compact stellar X-ray sources, pp
  215--265

\bibitem[\protect\citeauthoryear{{Corbel}, {Edwards}, {Tzioumis}, {Coriat},
  {Fender} \& {Brocksopp}}{{Corbel} et~al.}{2012}]{Corbel2012}
{Corbel} S.,  {Edwards} P.,  {Tzioumis} T.,  {Coriat} M.,  {Fender} R.,
  {Brocksopp} C.,  2012, The Astronomer's Telegram, 4410, 1

\bibitem[\protect\citeauthoryear{{Corral-Santana}, {Casares}, {Munoz-Darias},
  {Rodriguez-Gil}, {Shahbaz}, {Torres}, {Zurita} \& {Tyndall}}{{Corral-Santana}
  et~al.}{2013}]{Corral-Santana2013}
{Corral-Santana} J.~M.,  {Casares} J.,  {Munoz-Darias} T.,  {Rodriguez-Gil} P.,
   {Shahbaz} T.,  {Torres} M.~A.,  {Zurita} C.,    {Tyndall} A.~A.,  2013,
  Science, 339, 1048

\bibitem[\protect\citeauthoryear{{Cox}}{{Cox}}{2000}]{Cox2000}
{Cox} A.,  2000, Allen's astrophysical quantities, \skytel, 100, 72

\bibitem[\protect\citeauthoryear{{Cummings}, {Barthelmy}, {Baumgartner} \& {et
  al.}}{{Cummings} et~al.}{2012}]{Cummings2012b}
{Cummings} J.~R.,  {Barthelmy} S.~D.,  {Baumgartner} W.~H.,    {et al.} 2012,
  GRB Coordinates Network, 13775, 1

\bibitem[\protect\citeauthoryear{{Cummings}, {Gronwall}, {Grupe} \& {et
  al.}}{{Cummings} et~al.}{2012}]{Cummings2012}
{Cummings} J.~R.,  {Gronwall} C.,  {Grupe} D.,    {et al.} 2012, GRB
  Coordinates Network, 13774, 1

\bibitem[\protect\citeauthoryear{{de Ugarte Postigo}, {Sanchez-Ramirez},
  {Munoz-Darias}, {Gorosabel}, {Thoene} \& {Cabrera-Lavers}}{{de Ugarte
  Postigo} et~al.}{2012}]{deUgartePostigo2012}
{de Ugarte Postigo} A.,  {Sanchez-Ramirez} R.,  {Munoz-Darias} T.,  {Gorosabel}
  J.,  {Thoene} C.~C.,    {Cabrera-Lavers} A.,  2012, The Astronomer's
  Telegram, 4388, 1

\bibitem[\protect\citeauthoryear{{della Valle} \& {Benetti}}{{della Valle} \&
  {Benetti}}{1993}]{DellaValle1993}
{della Valle} M.,  {Benetti} S.,  1993, \iaucirc, 5890, 2

\bibitem[\protect\citeauthoryear{{Faulkner}, {Flannery} \& {Warner}}{{Faulkner}
  et~al.}{1972}]{Faulkner1972}
{Faulkner} J.,  {Flannery} B.~P.,    {Warner} B.,  1972, \apjl, 175, L79+

\bibitem[\protect\citeauthoryear{{Fender}}{{Fender}}{2006}]{Fender2006}
{Fender} R.,  2006, in Compact stellar X-ray sources, pp 381--419

\bibitem[\protect\citeauthoryear{{Fender}, {Homan} \& {Belloni}}{{Fender}
  et~al.}{2009}]{Fender2009}
{Fender} R.~P.,  {Homan} J.,    {Belloni} T.~M.,  2009, \mnras, 396, 1370

\bibitem[\protect\citeauthoryear{{Filippenko}, {Leonard}, {Matheson}, {Li},
  {Moran} \& {Riess}}{{Filippenko} et~al.}{1999}]{Filippenko1999}
{Filippenko} A.~V.,  {Leonard} D.~C.,  {Matheson} T.,  {Li} W.,  {Moran} E.~C.,
     {Riess} A.~G.,  1999, \pasp, 111, 969

\bibitem[\protect\citeauthoryear{{Harlaftis}, {Charles} \& {Horne}}{{Harlaftis}
  et~al.}{1997}]{Harlaftis1997}
{Harlaftis} E.~T.,  {Charles} P.~A.,    {Horne} K.,  1997, \mnras, 285, 673

\bibitem[\protect\citeauthoryear{{Horne} \& {Marsh}}{{Horne} \&
  {Marsh}}{1986}]{Horne1986}
{Horne} K.,  {Marsh} T.~R.,  1986, \mnras, 218, 761

\bibitem[\protect\citeauthoryear{{Hynes}, {Britt}, {Jonker}, {Wijnands} \&
  {Greiss}}{{Hynes} et~al.}{2012}]{Hynes2012}
{Hynes} R.~I.,  {Britt} C.~T.,  {Jonker} P.~G.,  {Wijnands} R.,    {Greiss} S.,
   2012, The Astronomer's Telegram, 4417, 1

\bibitem[\protect\citeauthoryear{{Hynes}, {Charles}, {Haswell}, {Casares},
  {Zurita} \& {Serra-Ricart}}{{Hynes} et~al.}{2001}]{Hynes2001}
{Hynes} R.~I.,  {Charles} P.~A.,  {Haswell} C.~A.,  {Casares} J.,  {Zurita} C.,
     {Serra-Ricart} M.,  2001, \mnras, 324, 180

\bibitem[\protect\citeauthoryear{{Hynes}, {Horne}, {O'Brien}, {Haswell},
  {Robinson}, {King}, {Charles} \& {Pearson}}{{Hynes} et~al.}{2006}]{Hynes2006}
{Hynes} R.~I.,  {Horne} K.,  {O'Brien} K.,  {Haswell} C.~A.,  {Robinson} E.~L.,
   {King} A.~R.,  {Charles} P.~A.,    {Pearson} K.~J.,  2006, \apj, 648, 1156

\bibitem[\protect\citeauthoryear{{Jordi}, {Grebel} \& {Ammon}}{{Jordi}
  et~al.}{2006}]{Jordi2006}
{Jordi} K.,  {Grebel} E.~K.,    {Ammon} K.,  2006, \aap, 460, 339

\bibitem[\protect\citeauthoryear{{King}}{{King}}{1993}]{King1993}
{King} A.~R.,  1993, \mnras, 260, L5+

\bibitem[\protect\citeauthoryear{{King}, {Kolb} \& {Burderi}}{{King}
  et~al.}{1996}]{King1996}
{King} A.~R.,  {Kolb} U.,    {Burderi} L.,  1996, \apjl, 464, L127+

\bibitem[\protect\citeauthoryear{{Maccarone}}{{Maccarone}}{2003}]{Maccarone200%
3}
{Maccarone} T.~J.,  2003, \aap, 409, 697

\bibitem[\protect\citeauthoryear{{Miller-Jones} \& {Sivakoff}}{{Miller-Jones}
  \& {Sivakoff}}{2012}]{Miller-Jones2012}
{Miller-Jones} J.~C.~A.,  {Sivakoff} G.~R.,  2012, The Astronomer's Telegram,
  4394, 1

\bibitem[\protect\citeauthoryear{{Mu{\~n}oz-Darias}, {Casares} \&
  {Mart{\'{\i}}nez-Pais}}{{Mu{\~n}oz-Darias} et~al.}{2008}]{Munoz-Darias2008}
{Mu{\~n}oz-Darias} T.,  {Casares} J.,    {Mart{\'{\i}}nez-Pais} I.~G.,  2008,
  \mnras, 385, 2205

\bibitem[\protect\citeauthoryear{{Mu{\~n}oz-Darias}, {Mart{\'{\i}}nez-Pais},
  {Casares}, {Dhillon}, {Marsh}, {Cornelisse}, {Steeghs} \&
  {Charles}}{{Mu{\~n}oz-Darias} et~al.}{2007}]{Munoz-Darias2007}
{Mu{\~n}oz-Darias} T.,  {Mart{\'{\i}}nez-Pais} I.~G.,  {Casares} J.,  {Dhillon}
  V.~S.,  {Marsh} T.~R.,  {Cornelisse} R.,  {Steeghs} D.,    {Charles} P.~A.,
  2007, \mnras, 379, 1637

\bibitem[\protect\citeauthoryear{{Orosz}, {Bailyn}, {Remillard}, {McClintock}
  \& {Foltz}}{{Orosz} et~al.}{1994}]{Orosz1994}
{Orosz} J.~A.,  {Bailyn} C.~D.,  {Remillard} R.~A.,  {McClintock} J.~E.,
  {Foltz} C.~B.,  1994, \apj, 436, 848

\bibitem[\protect\citeauthoryear{{Orosz}, {Remillard}, {Bailyn} \&
  {McClintock}}{{Orosz} et~al.}{1997}]{Orosz1997}
{Orosz} J.~A.,  {Remillard} R.~A.,  {Bailyn} C.~D.,    {McClintock} J.~E.,
  1997, \apjl, 478, L83

\bibitem[\protect\citeauthoryear{{{\"O}zel}, {Psaltis}, {Narayan} \&
  {McClintock}}{{{\"O}zel} et~al.}{2010}]{Ozel2010}
{{\"O}zel} F.,  {Psaltis} D.,  {Narayan} R.,    {McClintock} J.~E.,  2010,
  \apj, 725, 1918

\bibitem[\protect\citeauthoryear{{Pearson}, {Hynes}, {Steeghs}, {Jonker},
  {Haswell}, {King}, {O'Brien}, {Nelemans} \& {M{\'e}ndez}}{{Pearson}
  et~al.}{2006}]{Pearson2006}
{Pearson} K.~J.,  {Hynes} R.~I.,  {Steeghs} D.,  {Jonker} P.~G.,  {Haswell}
  C.~A.,  {King} A.~R.,  {O'Brien} K.,  {Nelemans} G.,    {M{\'e}ndez} M.,
  2006, \apj, 648, 1169

\bibitem[\protect\citeauthoryear{{Rau}, {Knust}, {Kann} \& {Greiner}}{{Rau}
  et~al.}{2012}]{Rau2012}
{Rau} A.,  {Knust} F.,  {Kann} D.~A.,    {Greiner} J.,  2012, The Astronomer's
  Telegram, 4380, 1

\bibitem[\protect\citeauthoryear{{Remillard} \& {McClintock}}{{Remillard} \&
  {McClintock}}{2006}]{Remillard2006b}
{Remillard} R.~A.,  {McClintock} J.~E.,  2006, \araa, 44, 49

\bibitem[\protect\citeauthoryear{{Russell}, {Fender}, {Hynes}, {Brocksopp},
  {Homan}, {Jonker} \& {Buxton}}{{Russell} et~al.}{2006}]{Russell2006}
{Russell} D.~M.,  {Fender} R.~P.,  {Hynes} R.~I.,  {Brocksopp} C.,  {Homan} J.,
   {Jonker} P.~G.,    {Buxton} M.~M.,  2006, \mnras, 371, 1334

\bibitem[\protect\citeauthoryear{{Russell}, {Fender} \& {Jonker}}{{Russell}
  et~al.}{2007}]{Russell2007}
{Russell} D.~M.,  {Fender} R.~P.,    {Jonker} P.~G.,  2007, \mnras, 379, 1108

\bibitem[\protect\citeauthoryear{{Russell}, {Lewis}, {Mundell} \& {et
  al.}}{{Russell} et~al.}{2012}]{Russell2012a}
{Russell} D.~M.,  {Lewis} F.,  {Mundell} C.~G.,    {et al.} 2012, The
  Astronomer's Telegram, 4456, 1

\bibitem[\protect\citeauthoryear{{Sbarufatti}, {Kennea}, {Burrows} \& {et
  al.}}{{Sbarufatti} et~al.}{2012}]{Sbarufatti2012}
{Sbarufatti} B.,  {Kennea} J.~A.,  {Burrows} D.~N.,    {et al.} 2012, The
  Astronomer's Telegram, 4383, 1

\bibitem[\protect\citeauthoryear{{Shahbaz}, {Charles} \& {King}}{{Shahbaz}
  et~al.}{1998}]{Shahbaz1998a}
{Shahbaz} T.,  {Charles} P.~A.,    {King} A.~R.,  1998, \mnras, 301, 382

\bibitem[\protect\citeauthoryear{{Shahbaz} \& {Kuulkers}}{{Shahbaz} \&
  {Kuulkers}}{1998}]{Shahbaz1998}
{Shahbaz} T.,  {Kuulkers} E.,  1998, \mnras, 295, L1

\bibitem[\protect\citeauthoryear{{Shahbaz}, {Naylor} \& {Charles}}{{Shahbaz}
  et~al.}{1993}]{Shahbaz1993}
{Shahbaz} T.,  {Naylor} T.,    {Charles} P.~A.,  1993, \mnras, 265, 655

\bibitem[\protect\citeauthoryear{{Shahbaz}, {van der Hooft}, {Charles},
  {Casares} \& {van Paradijs}}{{Shahbaz} et~al.}{1996}]{Shahbaz1996}
{Shahbaz} T.,  {van der Hooft} F.,  {Charles} P.~A.,  {Casares} J.,    {van
  Paradijs} J.,  1996, \mnras, 282, L47

\bibitem[\protect\citeauthoryear{{Shakura} \& {Sunyaev}}{{Shakura} \&
  {Sunyaev}}{1973}]{Shakura1973}
{Shakura} N.~I.,  {Sunyaev} R.~A.,  1973, \aap, 24, 337

\bibitem[\protect\citeauthoryear{{Smak}}{{Smak}}{1969}]{Smak1969}
{Smak} J.,  1969, \actaa, 19, 155

\bibitem[\protect\citeauthoryear{{Tomsick}, {DelSanto} \& {Belloni}}{{Tomsick}
  et~al.}{2012}]{Tomsick2012}
{Tomsick} J.~A.,  {DelSanto} M.,    {Belloni} T.,  2012, The Astronomer's
  Telegram, 4393, 1

\bibitem[\protect\citeauthoryear{{Torres}, {Callanan}, {Garcia}, {Zhao},
  {Laycock} \& {Kong}}{{Torres} et~al.}{2004}]{Torres2004}
{Torres} M.~A.~P.,  {Callanan} P.~J.,  {Garcia} M.~R.,  {Zhao} P.,  {Laycock}
  S.,    {Kong} A.~K.~H.,  2004, \apj, 612, 1026

\bibitem[\protect\citeauthoryear{{Torres}, {Casares}, {Mart{\'{\i}}nez-Pais} \&
  {Charles}}{{Torres} et~al.}{2002}]{Torres2002}
{Torres} M.~A.~P.,  {Casares} J.,  {Mart{\'{\i}}nez-Pais} I.~G.,    {Charles}
  P.~A.,  2002, \mnras, 334, 233

\bibitem[\protect\citeauthoryear{{van der Klis}}{{van der
  Klis}}{2006}]{vanderklis2006}
{van der Klis} M.,  2006, pp 39--112

\bibitem[\protect\citeauthoryear{{van Paradijs} \& {McClintock}}{{van Paradijs}
  \& {McClintock}}{1994}]{vanParadijs1994}
{van Paradijs} J.,  {McClintock} J.~E.,  1994, \aap, 290, 133

\bibitem[\protect\citeauthoryear{{Zurita}, {Casares}, {Shahbaz}, {Charles},
  {Hynes}, {Shugarov}, {Goransky}, {Pavlenko} \& {Kuznetsova}}{{Zurita}
  et~al.}{2000}]{Zurita2000}
{Zurita} C.,  {Casares} J.,  {Shahbaz} T.,  {Charles} P.~A.,  {Hynes} R.~I.,
  {Shugarov} S.,  {Goransky} V.,  {Pavlenko} E.~P.,    {Kuznetsova} Y.,  2000,
  \mnras, 316, 137

\bibitem[\protect\citeauthoryear{{Zurita}, {Torres}, {Steeghs},
  {Rodr{\'{\i}}guez-Gil}, {Mu{\~n}oz-Darias}, {Casares}, {Shahbaz},
  {Mart{\'{\i}}nez-Pais}, {Zhao}, {Garcia}, {Piccioni} \& {et al.}}{{Zurita}
  et~al.}{2006}]{Zurita2006}
{Zurita} C.,  {Torres} M.~A.~P.,  {Steeghs} D.,  {Rodr{\'{\i}}guez-Gil} P.,
  {Mu{\~n}oz-Darias} T.,  {Casares} J.,  {Shahbaz} T.,  {Mart{\'{\i}}nez-Pais}
  I.~G.,  {Zhao} P.,  {Garcia} M.~R.,  {Piccioni} A.,    {et al.} 2006, \apj,
  644, 432

\end{thebibliography}
\nocite{Cummings2012}
\nocite{Cummings2012b}
\nocite{Russell2007}

\label{lastpage}
\end{document}